\documentclass{svproc}
\usepackage{xcolor}
\usepackage{amsmath,amssymb, bbm}
\usepackage{epsfig}

\usepackage{graphicx,float}
\usepackage{subcaption}
\usepackage{url}

\date{}
\begin{document}

\mainmatter

\title{Structural Change and Random Graph Models in Global Oil Trade Networks}

\author{Anthony Bonato\inst{1}\thanks{Supported by an NSERC Discovery Grant.} \and Vincent Luong\inst{1} \and Kyne Santos\inst{1}}

\institute{Toronto Metropolitan University, Toronto, Ontario, Canada}

\maketitle

\begin{abstract}
We studied structural change in global oil trade using a network approach. Using UN Comtrade data, we examined the temporal evolution of international trade networks, with an emphasis on crude oil. Weighted in-degree identified major changes in country rankings in 1991, 2011, 2017, and 2021, while PageRank detected pronounced changes around 1991 and 2024. The Louvain algorithm identified clear geographic communities within the overall trade network. In the oil trade network, modularity declined from the 1990s to the 2010s, with node2vec embeddings showing weaker clustering in 2011 than in 1991. We also compared the oil trade network with several random graph models using 3- and 4-node subgraph profiles and machine learning classification. The oil trade network was consistently classified as a Chung-Lu graph, while the Geometric model was not favored, suggesting that a model based on the degree distribution better matched its subgraph profiles than the other models considered.
\end{abstract}

\section{Introduction}\label{intro}

International trade forms a large, evolving network whose structure reflects economic activity, geography, and geopolitical relations. Its organization changes over time as economies develop and trading relationships shift \cite{fagiolo,kosztyan}. Geopolitical conflicts and sanctions may also alter established trade patterns \cite{telarico}. Oil trade provides a particularly useful setting for studying such changes. Crude oil production is concentrated in a relatively small number of countries, while demand is global, creating trade patterns that need not align with geographic proximity. A network approach allows changes in individual countries' roles to be studied alongside changes in community structure and connectivity across the trade network.

The world trade network has been studied from several network-science perspectives. Early work identified scale-free structure in international trade \cite{serrano}. The temporal evolution of the network and its structural properties were studied in \cite{fagiolo}, while connections between network structure and national economic characteristics were examined in \cite{garlaschelli2004,garlaschelli2006}. The emergence of major intermediary trade hubs was examined in \cite{amador}, and the hierarchical structure of world commodity trade networks was studied in \cite{nobi}. The response of trade networks to economic crises was examined in \cite{kosztyan}, while the effects of sanctions on international trading relationships were studied in \cite{telarico}. 

We study structural change in global trade networks using United Nations Comtrade data, with particular emphasis on crude oil trade. We use centrality measures, community detection, network embeddings, and model selection techniques with random graph models. Our analysis identified both short-term disruptions and longer-term structural changes in global oil trade. Changes in country rankings coincided with several major geopolitical and economic events, while community detection and network embeddings indicated weaker community structure in oil trade networks (more specifically, in what we call the K-3 Oil Trade Network defined in Section~3) over time. Finally, comparison with random graph models showed that degree-based models better reproduced the subgraph structure of the time-aggregated K-3 Oil Trade Network among the models considered.

Section~2 reviews previous work on international trade networks and network methods for detecting structural change. Section~3 describes the UN Comtrade data and the methods used in the analysis, including centrality measures, the Louvain algorithm, node embeddings, and graph models. The corresponding results are presented in Section~4. The concluding section summarizes the main findings and discusses several directions for further study.

Additional background on graph theory can be found in \cite{west}, and further background on complex networks is available in the book \cite{bonato}.

\section{Literature Review}\label{aml}

One of the earliest empirical studies of the world trade network was conducted in 2003 using UN Comtrade data from 2000 \cite{serrano}. The study reported scale-free structure in the network. The evolution of network statistics (such as connectivity, assortativity, and centrality) over 20 years (between 1981 and 2000) was studied in \cite{fagiolo}. Work by Garlaschelli found a significant correlation between a country's GDP and its position in the world trade network \cite{garlaschelli2004,garlaschelli2006}. Amador et al.\ identified major structural changes in global value chains, including the rise of China and distinct central roles for Germany, the USA, China, and Japan \cite{amador}. Nobi et al.\ found an increase in the hierarchical organization of several world commodity trade networks and identified China's growing role as an important contributor to this structural evolution \cite{nobi}. For binary international trade networks, Squartini et al.\ showed that higher-order properties can largely be traced to their degree sequences \cite{squartini1}.

Nobi et al.\ found that the general increase in hierarchy from 1995 to 2013 was interrupted during periods of financial crisis \cite{nobi}. Interest in the resilience of trade networks to shocks has grown following events such as the 2008 financial crisis and the COVID-19 pandemic. A study of network statistics, cluster analysis, and causality in this context appears in \cite{kosztyan}. Geopolitical conflicts and sanctions may also reshape international trade. A network study of sanctions on Russia following its 2014 invasion of Crimea appears in \cite{telarico}, where stochastic block modeling revealed a rearrangement of Russian trade flows away from sanctioning states without necessarily diminishing Russia's role in the network.

Several studies have focused specifically on the structure and evolution of oil trade networks. Fracasso et al.\ \cite{fracasso} analyzed changes in centrality and community structure from 1995 to 2014, while Xie et al.\ \cite{xieoil} studied the evolving robustness of global oil trade networks from 1988 to 2017. More recent work has examined the relationship between geopolitical risk and network structure \cite{tao}, as well as the spatiotemporal evolution of crude oil trade through 2023 \cite{xu}. In contrast to these studies, we also compare the oil trade network with several random graph models using subgraph profiles and machine learning classification.

Network embeddings provide an approach to community detection by capturing structural similarities between nodes that need not be adjacent; see \cite{hou} for an overview and \cite{node2vec} for node2vec. Related work on global banking networks used PageRank and other centrality measures to track changes associated with the 2008 financial crisis \cite{rabobank}.

\section{Methods and Data}\label{methods}

We used data from the United Nations Comtrade (UN Comtrade) database \cite{comtrade}. UN Comtrade is a database that provides a breakdown of a country's imports and exports over many years. The data contains the U.S. dollar value traded between countries and is broken down by the type of good traded. The data is collected by the United Nations Statistics Division, then processed and made available through UN Comtrade. The data can be accessed freely at \url{https://comtrade.un.org/}.

Using UN Comtrade data, we then created a network in which each node represents a unique country. For each year, we created an overall trade network with a directed edge from country A to country B, with weight equal to the total exports from country A to country B in U.S.\ dollar terms.

We created an annual oil trade network based solely on crude oil exports and imports, rather than total trade. We used this network to examine periods of disruption in global oil trade. For the oil trade network, we filtered the trade data using the Harmonized System (HS) code 2709.

We also constructed the \textit{K-3 Oil Trade Network}. For each year, this network had the same nodes as the oil trade network, but for each country, we retained only its top three export and top three import partners. The resulting edges were treated as unweighted and undirected. This produced a sparse backbone of countries' strongest trading relationships suitable for the graph-model analysis in Section~\ref{methods_models}. For analyses spanning multiple years, we took the union of the corresponding annual K-3 networks. Our codebase, including the data, is available at \url{https://github.com/vinceLuong/tradeNetwork}.

\subsection{Network Centrality}

We used weighted degree centrality and PageRank to compare countries' roles in trade networks. For a weighted directed network, the weighted out-degree of a country is the total weight of its outgoing edges, while its weighted in-degree is the total weight of its incoming edges. Since edge weights represent trade values in U.S. dollars, these measures provide information about the value of trade associated with each country.

For the exporter analysis, we reversed the direction of the trade edges so that an edge from country $u$ to country $v$ represents oil imported by $u$ from $v$. In this reverse-edge network, the weighted in-degree of a country is therefore its total export value. We used this orientation for both weighted in-degree and PageRank so that larger centrality values correspond to countries that play more prominent roles as oil exporters.

PageRank measures the probability that a random walk with teleportation visits a given node \cite{pagerank}. In this random walk, the probability of traversing an edge is proportional to its weight. The PageRank random walk may also teleport rather than traverse an edge. We used a teleportation probability of 15\%, with the teleportation destination chosen uniformly at random. PageRank typically favors nodes with high in-degree, since this correlates with a higher probability that their in-edges are used.

\subsection{Louvain Algorithm}

For the community analysis, we treated the trade network as an undirected weighted network, with the weight of an edge between two countries given by their total bilateral trade. We used the Louvain algorithm \cite{louvain} to identify communities in this network. The algorithm seeks a partition with high \textit{modularity}, so that the weight of edges within communities is large relative to that expected under a null model. The modularity of a partition is defined by
$$
\frac{1}{2m}\sum_{i,j}\left(A_{ij}-\frac{k_i k_j}{2m}\right)\delta(c_i,c_j),
$$
where $A_{ij}$ is the edge weight of edge $ij$, $k_i$ is the sum of edge weights of edges incident to $i$, $m$ is the sum of all edge weights in the network, and $\delta(c_i, c_j)$ is 1 if nodes $i$ and $j$ are in the same community, and 0 otherwise.

Louvain starts with each node in its own community and alternates between local node moves that increase modularity and community aggregation. In the local-moving phase, a node is moved to a neighboring community when doing so increases modularity. The resulting communities are then contracted into single nodes, and the process is repeated until no further improvement is obtained.

\subsection{Node Embeddings}

Node embeddings represent nodes in a network as vectors in a Euclidean space. We used node2vec \cite{node2vec} to construct embeddings of the K-3 Oil Trade Network. The algorithm learns these representations from biased second-order random walks that balance local and outward exploration of the network.

Suppose that a walk has moved from node $t$ to node $v$, and let $x$ be adjacent to $v$. Let $d_{tx}$ denote the distance between $t$ and $x$, and let $w_{vx}$ denote the weight of the edge $vx$, which is equal to 1 in the unweighted K-3 Oil Trade Network. The unnormalized transition weight from $v$ to $x$ is
\[
\pi_{vx}=\alpha_{pq}(t,x)w_{vx},
\]
where
\[
\alpha_{pq}(t,x)=
\begin{cases}
\frac{1}{p}, & \text{if } d_{tx}=0,\\
1, & \text{if } d_{tx}=1,\\
\frac{1}{q}, & \text{if } d_{tx}=2.
\end{cases}
\]
The transition probabilities are obtained by normalizing these weights over the neighbors of $v$. The parameter $p$ controls the likelihood of immediately returning to the previous node, while $q$ controls the tendency of the walk toward local or more outward exploration.

Once the walks are generated, node2vec uses the Skip-gram model to learn the embeddings. The walks form a corpus analogous to sentences, with nodes playing the role of words. Node2vec learns the embeddings by predicting nearby nodes in the walks from a given node. More information on the Skip-gram model and word2vec can be found in \cite{word2vec}.

Typically, node2vec embeddings are high-dimensional to capture more information about nodes' neighborhoods. In our analysis of trade networks, we used node2vec to produce embeddings in $\mathbb{R}^{128}$, and then used UMAP \cite{umap} to reduce the embeddings to two dimensions. We examined the resulting two-dimensional plots to identify community structure, similar to that identified by the Louvain algorithm.

\subsection{Graph Models}\label{methods_models}

We compared the K-3 Oil Trade Network with five random graph models having different structural properties. In our analysis, we considered the \textit{Erdős--Rényi} (ER), \textit{Chung--Lu}, \textit{Preferential Attachment}, \textit{Configuration}, and \textit{Geometric} models. In the ER model, each possible edge is included independently with a fixed probability. The Chung--Lu model instead uses edge probabilities based on the expected degrees of the two endpoints. In the Preferential Attachment model, nodes are added over discrete time steps, with new nodes more likely to connect to higher-degree nodes. The Configuration model randomly generates graphs from a specified degree sequence. For further details on these four models, see \cite{character}. In the Geometric model \cite{penrose}, nodes are represented by randomly distributed points in a hypercube of given dimension, with two nodes adjacent whenever the distance between their corresponding points is at most a specified threshold.

Our goal was to determine which of these graph models most closely resembles the trade network based on small-subgraph counts. Following the approach in \cite{character}, we used counts of induced subgraphs on three and four nodes. Up to isomorphism, there are four graphs on three nodes and eleven graphs on four nodes. Their counts form a combined 3- and 4-profile, represented by a vector in $\mathbb{R}^{15}$.

Using the combined 3- and 4-profile as a feature vector, we applied several machine learning algorithms to classify the trade network according to the graph models considered. For each graph model, we generated 100 random graphs with the same number of nodes and edges as the corresponding K-3 Oil Trade Network, with model-specific parameters chosen to match these quantities. We considered support vector machines (SVMs), decision trees, random forests, and AdaBoost, with model parameters selected via a grid search with 5-fold cross-validation. We applied this model selection framework to the K-3 Oil Trade Network for both individual years and networks obtained by combining data across multiple years. We focused on the K-3 Oil Trade Network since it is unweighted and undirected, consistent with the graphs generated by the models considered.

\section{Results}\label{results}

\subsection{Centrality Measures}

\begin{figure}[h]
    \centering
    \includegraphics[width=0.6\linewidth]{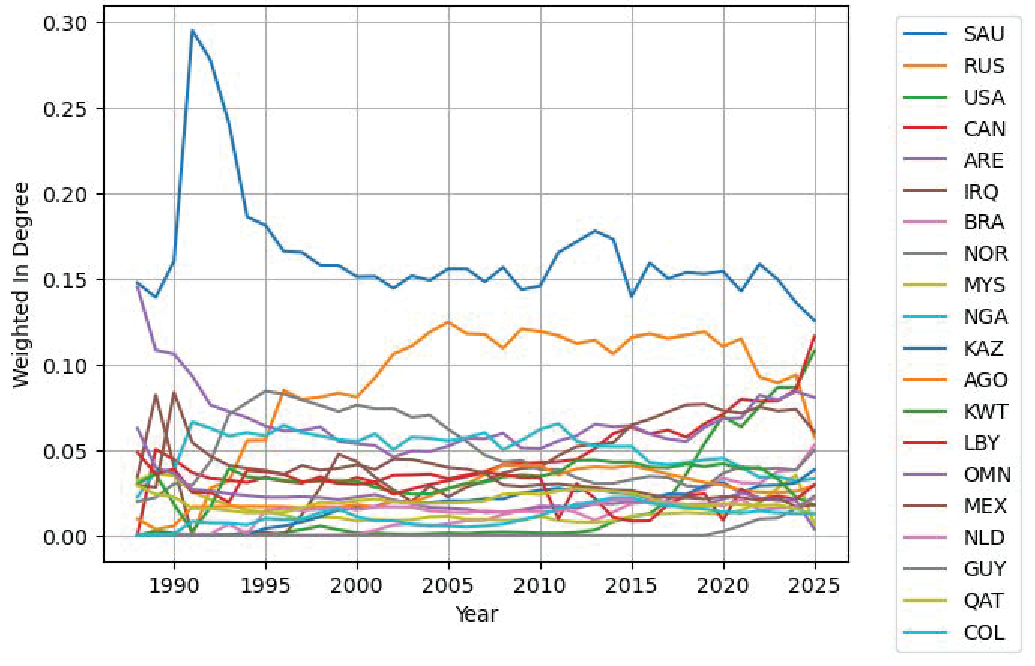}
    \caption{Weighted in-degree scores among the top 20 oil exporters.}
    \label{fig:in-degree}
\end{figure}

On the reverse-edge network, a weighted edge from country $u$ to $v$ contains the total value (in U.S.\ dollars) of crude petroleum oil imported by $u$ from $v$ in a given year. The weighted in-degree of a country thus gives its total export value in a year, which we normalize to a centrality score between 0 and 1. We used the 20 countries with the highest weighted in-degree scores in 2024 as a fixed cohort for the time series analysis. Figure~\ref{fig:in-degree} displays the dynamics among the top 20 from 1988 to 2025. Saudi Arabia remained the dominant oil exporter, with Russia closely following.

\begin{figure}[htpb!]
\centering
\begin{minipage}{0.48\textwidth}
    \centering
    \includegraphics[width=\linewidth]{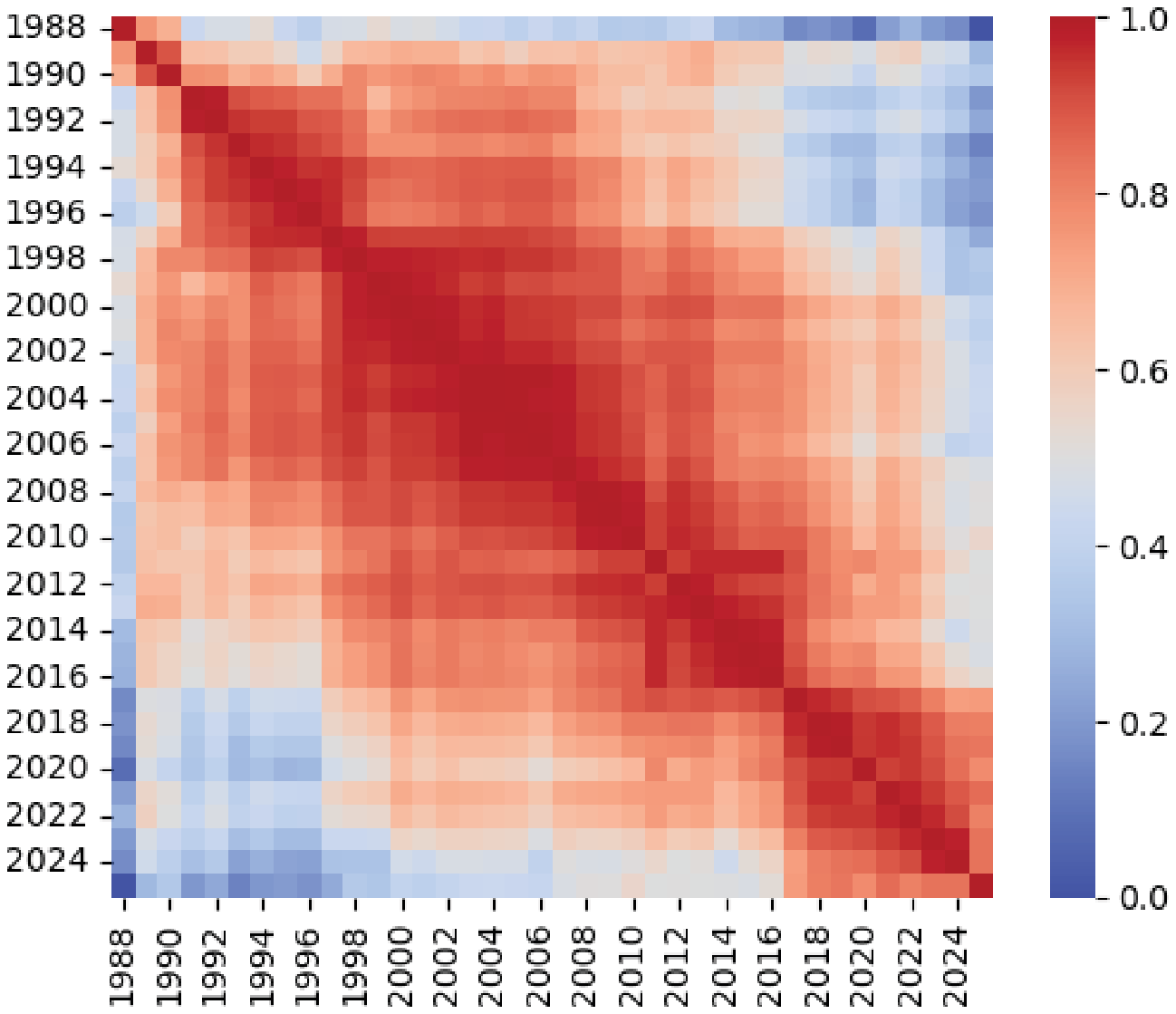}
\end{minipage}
\hfill
\begin{minipage}{0.48\textwidth}
    \centering
    \includegraphics[width=\linewidth]{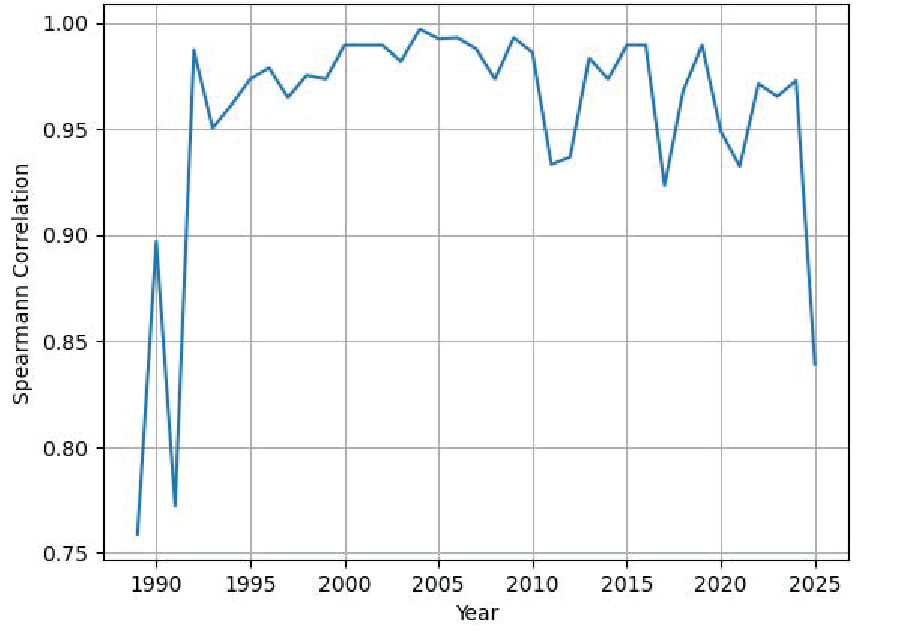}
\end{minipage}
\caption{Spearman's rank correlation of weighted in-degree rankings. Left: heatmap across years. Right: year-to-year rank correlation.}
\label{fig:WIDcorrelation}
\end{figure}

To identify years of substantial rank rearrangement within this cohort, we computed Spearman's rank correlation coefficient between their rankings in two given years (Figure~\ref{fig:WIDcorrelation}). Light-colored squares near the main diagonal indicate a clear rearrangement of rankings over a short period. By plotting Spearman's rank correlation coefficient across one-year time windows (Figure~\ref{fig:WIDcorrelation}), we see that the years 1991, 2011, 2017, and 2021 all exhibit stark rearrangements. 

\begin{figure}[htpb!]
\centering
\begin{minipage}{0.5\textwidth}
    \centering
    \includegraphics[width=\linewidth]{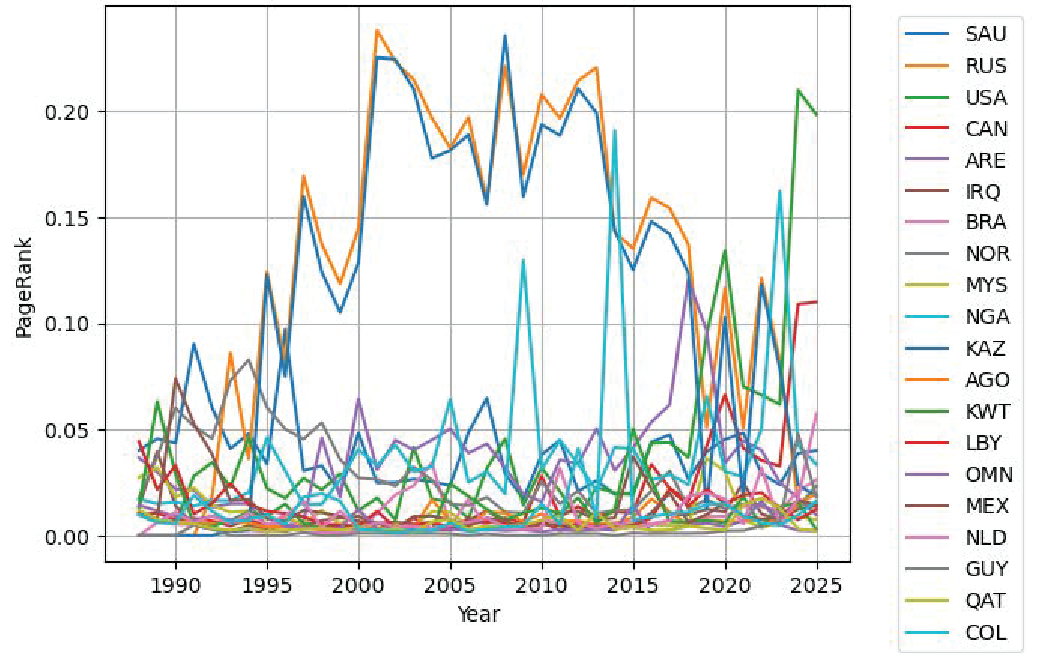}
\end{minipage}
\hfill
\begin{minipage}{0.48\textwidth}
    \centering
    \includegraphics[width=\linewidth]{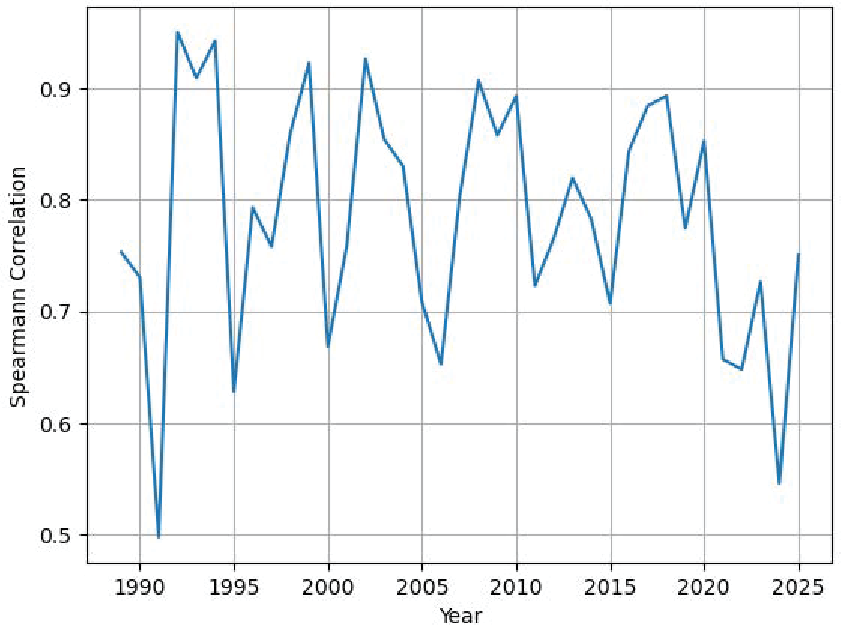}
\end{minipage}
\caption{PageRank analysis of the top 20 oil exporters. Left: PageRank scores across years. Right: year-to-year Spearman's rank correlation of PageRank rankings.}
\label{fig:pagerank}
\end{figure}

We also used PageRank, rather than weighted in-degree, to rank the top 20 exporters. The PageRank rankings exhibit different temporal dynamics from those based on weighted in-degree, with the most pronounced year-to-year changes occurring around 1991 and 2024 (Figure~\ref{fig:pagerank}).

Several of these changes coincided with major disruptions in global oil markets. The change in 1991 occurred during the Gulf War and also coincided with changes in country reporting following the dissolution of the Soviet Union, while the change in 2011 coincided with the Arab Spring and the Libyan Civil War. The change in 2017 occurred during a period of shifting global production and rapid growth in U.S.\ crude oil exports, while 2021 followed the major disruption to oil demand caused by the COVID-19 pandemic. 

\subsection{Louvain Algorithm}

\begin{figure}
    \centering
    \includegraphics[width=0.8\linewidth]{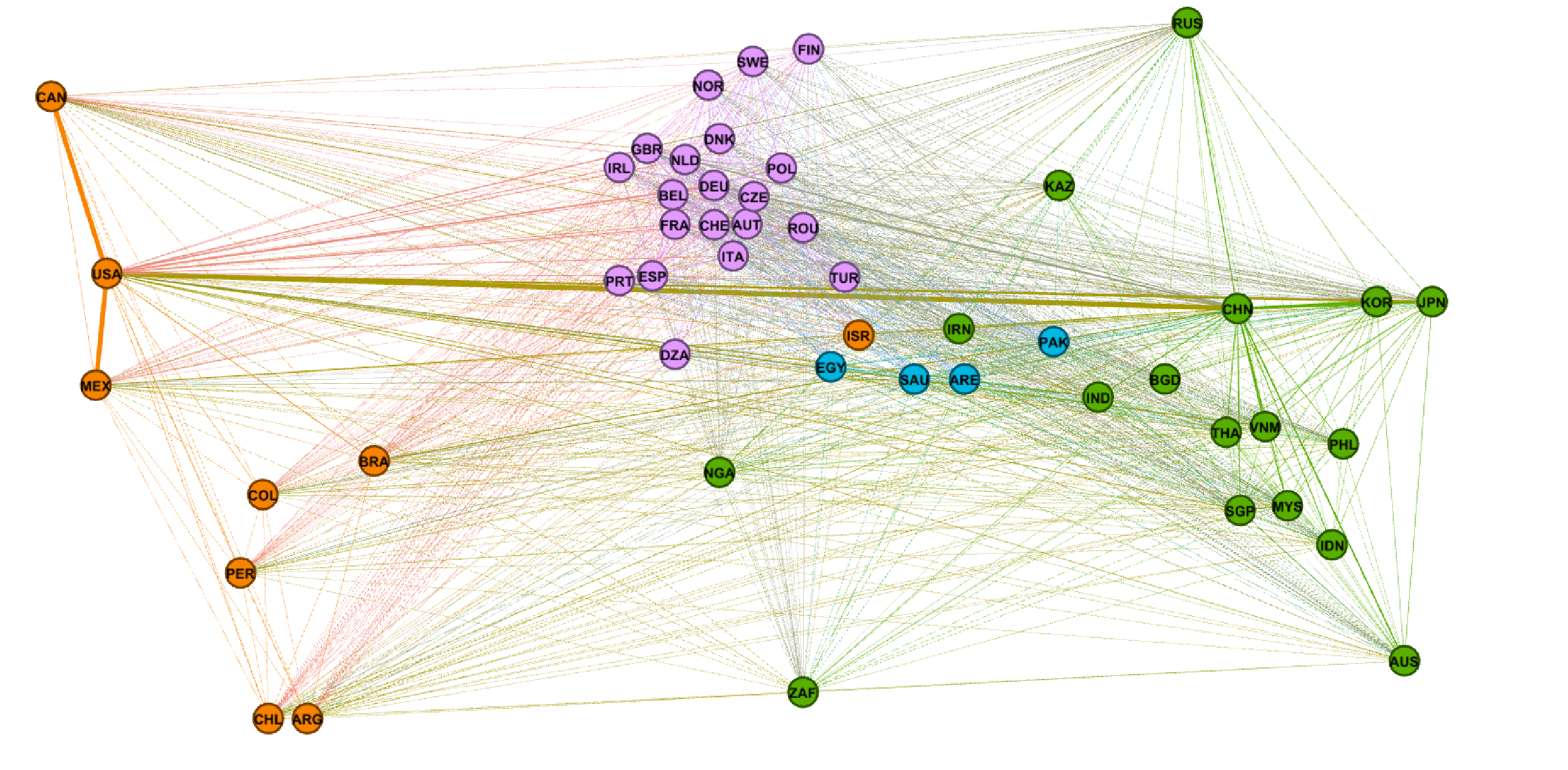}
    \caption{Communities identified by Louvain in the 2020 G50 trade network. Countries with the same color are in the same community. Nodes are placed based on their geographical location. }
    \label{fig:louvain}
\end{figure}

We first examined community structure in the overall trade network using the Louvain algorithm. We found that the communities formed around different geographic locations. Figure~\ref{fig:louvain} shows the communities obtained by applying the Louvain algorithm to the overall trade network from 2020. The figure was made using Gephi \cite{gephi}. We chose to apply the Louvain algorithm only to the top 50 countries by GDP (G50) because of their significance in the global economy. The country nodes are placed based on each country's latitude and longitude. The orange community is mainly in the Americas; the purple community is mainly in Western Europe; the green community contains countries from Eastern Europe, Asia, Oceania, and Africa; and the blue community is mainly in the Middle East. The geographic concentration of these communities is consistent with the role of proximity and transportation costs in international trade.

\subsection{Node Embeddings}
\begin{figure}[htpb!]
    \centering
    \includegraphics[width=0.5\linewidth]{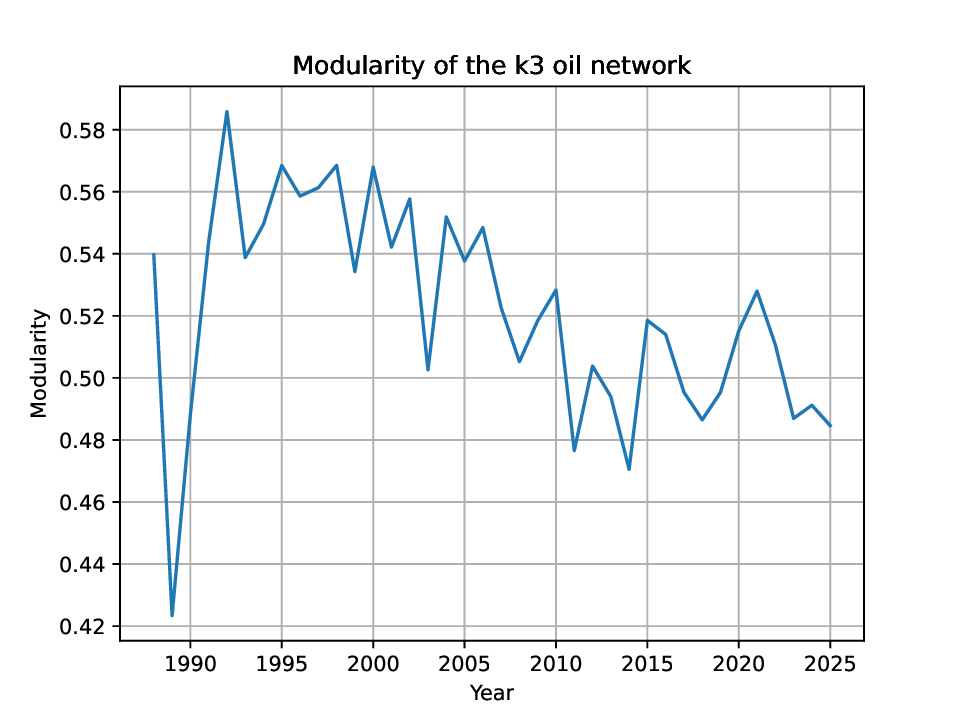}
    \caption{Modularity scores in the K-3 Oil Trade Network.}
    \label{fig:modularity}
\end{figure}

\begin{figure}[htpb!]
\centering
\begin{minipage}{0.48\textwidth}
    \centering
    \includegraphics[width=\linewidth]{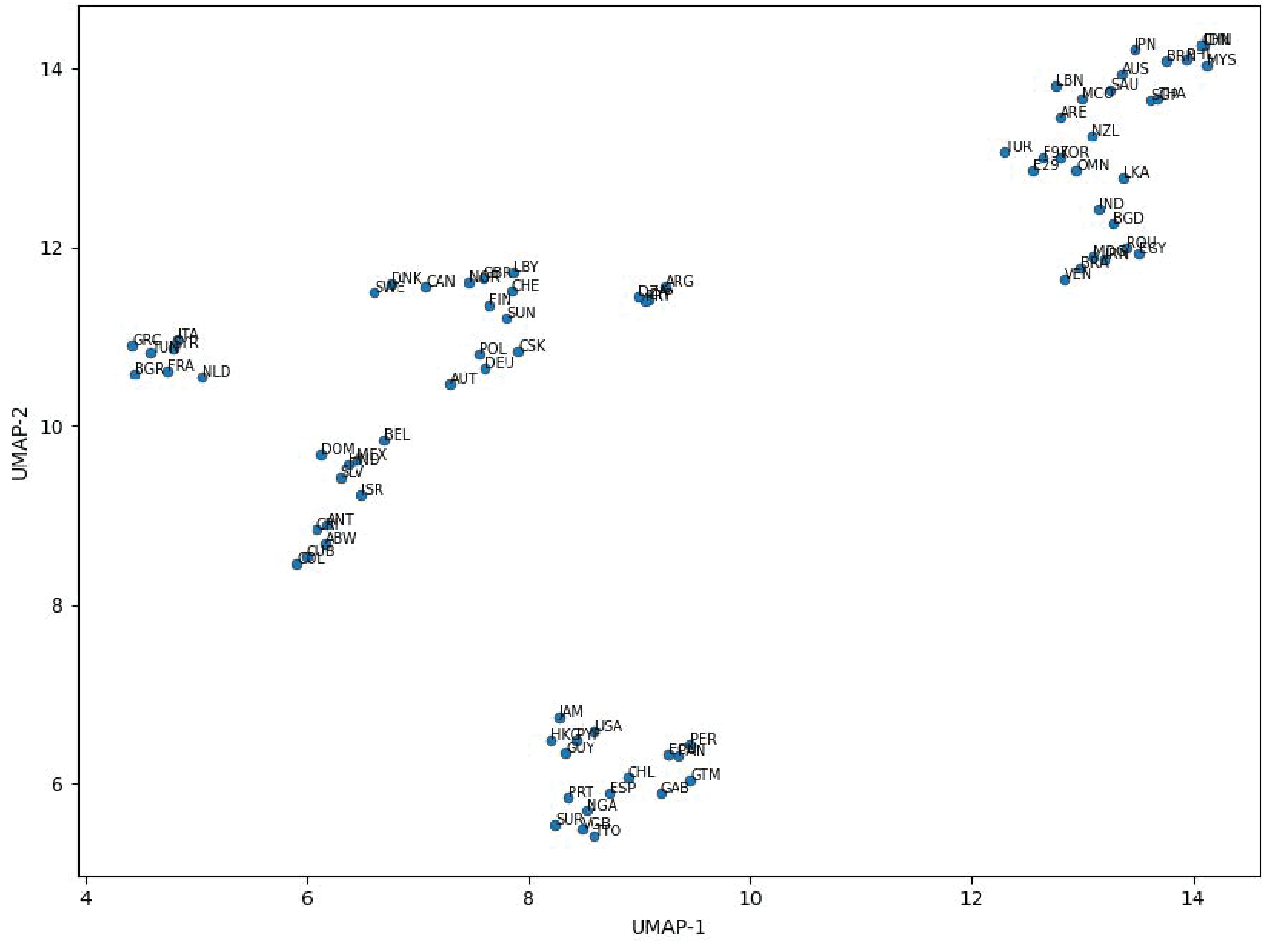}
\end{minipage}
\hfill
\begin{minipage}{0.48\textwidth}
    \centering
    \includegraphics[width=\linewidth]{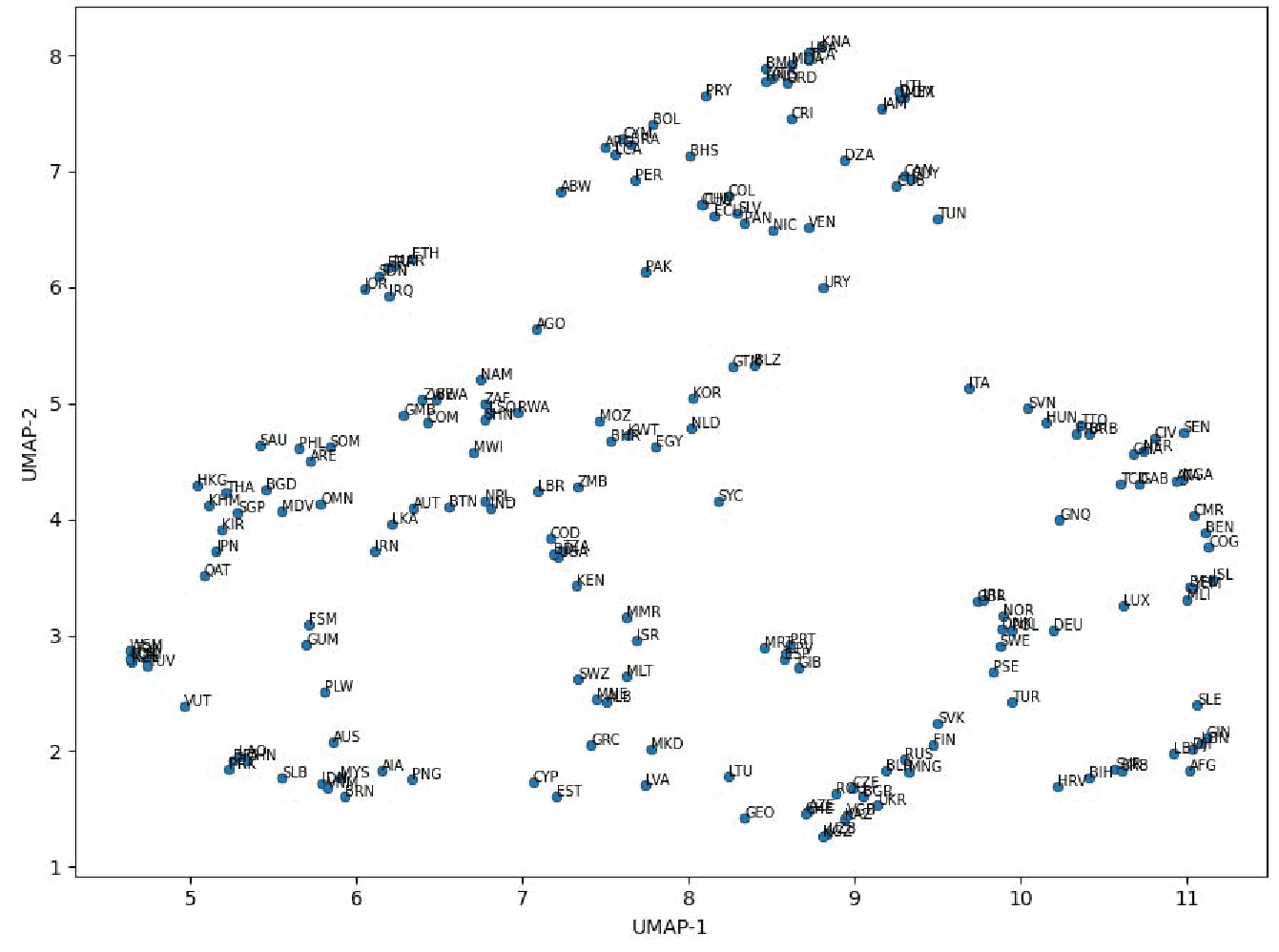}
\end{minipage}
\caption{UMAP projections of node2vec embeddings of the K-3 Oil Trade Network. Left: 1991. Right: 2011. The more pronounced clustering in 1991 is consistent with the higher modularity observed in the earlier period. Examples of clusters appearing in 1991 include Australia, Saudi Arabia, and Oman (top right); the USA, Spain, and Chile (bottom); and Italy, France, and the Netherlands (left).}
\label{fig:umap}
\end{figure}

% less community structure in later years
% Run louvain on a structured umap year to see if communities are the same
% CAN, USA become big exporters, could influence why everything is so spread apart
% Group years together for trend looking
We next examined community structure in the K-3 Oil Trade Network. Modularity declined overall from the 1990s to the 2010s (Figure~\ref{fig:modularity}), consistent with weaker community structure in later years. 
We also generated UMAP projections of node2vec embeddings for 1991 and 2011 (Figure~\ref{fig:umap}); the 1991 projection exhibits more pronounced clustering. The projections are consistent with the modularity trend.

\subsection{Graph Models}

\begin{figure}[htpb!]
    \centering
    \includegraphics[width=0.6\linewidth]{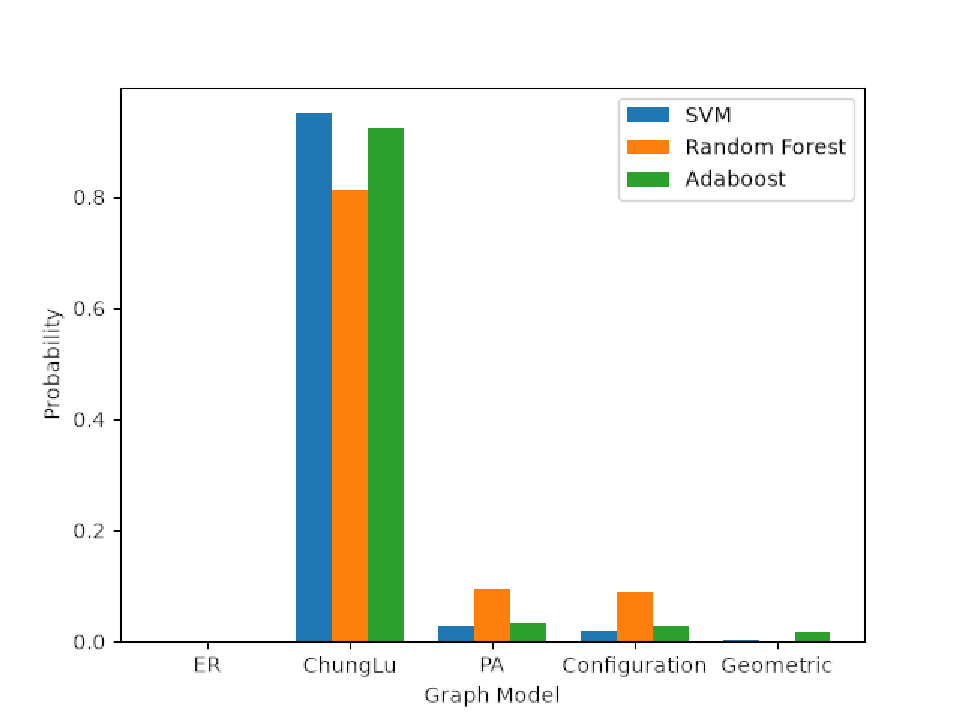}
    \caption{Machine learning classifier outputs for the time-aggregated K-3 Oil Trade Network. The probability-based classifiers strongly favor the Chung-Lu class.}
    \label{fig:ml_historic}
\end{figure}

In the K-3 Oil Trade Network, we have data spanning the years 1988 to 2025. We first analyzed the time-aggregated K-3 network obtained by taking the union of the annual networks from 1988 to 2025, yielding 236 nodes and 2,586 edges. After we trained the machine learning algorithms on 100 generated graphs from each graph model, each algorithm classified the time-aggregated K-3 network as a Chung-Lu graph. The probability-based classifiers strongly favored the Chung-Lu class, as shown in Figure~\ref{fig:ml_historic}. The decision tree model is omitted from the figure since only its predicted class was recorded.

The Chung-Lu classification points to degree heterogeneity as an important feature of the K-3 network. The Geometric model was not favored despite the geographic nature of international trade. However, the Geometric model does not use actual country locations or other factors captured by economic gravity models. Thus, the result shows only that, among the models considered, the degree-based models better reproduce the observed subgraph profiles.

We next analyzed each year's K-3 Oil Trade Network individually. The machine learning models typically classified the annual networks into either the Configuration or Chung-Lu class. Both models are based on the observed degree sequence.  The Configuration model uses that sequence directly, while the Chung-Lu model matches the observed degrees only in expectation, so generated Chung-Lu graphs may have degree sequences that differ from the observed sequence. This variation may explain why the classifiers favor the Configuration model in some years, particularly for the smaller annual networks. The annual results broadly agree with those for the time-aggregated K-3 network. Among the candidate models, degree-based models more closely reproduced the subgraph features used by the classifiers than the Geometric model.

\section{Discussion and Future Directions}

The results identified both short-term disruptions and longer-term changes in the structure of global oil trade. Weighted in-degree rankings showed substantial year-to-year rearrangements in 1991, 2011, 2017, and 2021. These dates coincided with major changes in global oil markets, including those associated with the Gulf War, the Arab Spring, and the Libyan Civil War, shifts in global oil production and U.S.\ exports, and the recovery from the disruption caused by the COVID-19 pandemic. PageRank also detected pronounced changes around 1991 and 2024. The overall trade network exhibited clear geographic communities under the Louvain algorithm, whereas modularity in the K-3 Oil Trade Network declined from the 1990s to the 2010s. The node2vec embeddings were consistent with this pattern, with stronger clustering in 1991 than in 2011. Finally, the K-3 Oil Trade Network was consistently classified as a Chung-Lu graph or a Configuration graph by the machine learning models, while the Geometric model was not favored.

Periods of economic and geopolitical disruption coincide with substantial changes in the relative positions of major trading countries. Over longer time scales, the K-3 Oil Trade Network shows weaker community structure in the 2010s than in the 1990s, suggesting that major oil-trading relationships became less partitioned into distinct groups over time. All classifiers identified the time-aggregated K-3 network as a Chung-Lu graph, while the Geometric model performed poorly, indicating that degree-based models better reproduced its subgraph structure among the models considered. One interpretation is that oil trade is organized less by geography than by the concentration of trade around major exporters and importers.

A next step is to distinguish structural changes associated with geopolitical conflicts from those driven by broader economic factors. The model selection approach could be extended to other spatial or gravity-based network models, and the effects of trade policy shocks, such as tariffs and sanctions, merit separate study. Comparisons with commodities such as steel and aluminum, semiconductors, and agricultural products could help determine whether the decline in community structure observed in the K-3 Oil Trade Network is more broadly evident.

\end{document}